\documentclass[printer]{aa}
\def\gsim{ \lower .75ex \hbox{$\sim$} \llap{\raise .27ex \hbox{$>$}} }
\def\lsim{ \lower .75ex\hbox{$\sim$} \llap{\raise .27ex \hbox{$<$}} }
\usepackage{graphicx}

\authorrunning{Guetta \& Piran}
\titlerunning{Short duration GRBs}

\begin{document}
\title{The luminosity and redshift distributions of short-duration GRBs}
\author{Dafne Guetta\inst{1,2}
\and Tsvi Piran\inst{2}}

\institute{ Osservatorio astronomico of Rome v. Frascati 33 00040
Monte Porzio Catone, Italy  \and  Racah Institute for Physics, The
Hebrew University, Jerusalem 91904, Israel}
\date{Received July 21 2004; accepted December 22, 2004}

\abstract{ Using the BATSE peak flux distribution we rederive the
short GRBs luminosity function and compare it with the observed
redshift distribution of long bursts. We show that both
distributions are compatible with the assumption that short  as
well as long bursts follow the star formation rate. In this case
the difference between the two observed distributions can be
interpreted as arising mostly from differences in the detector's
sensitivity to long and short bursts, while the local rate of
short bursts is $0.1 h_{70}^3$Gpc$^{-3}$yr$^{-1}$ . We also
consider the possibility that short GRBs may be associated with
binary neutron star mergers and estimate the effect of the merging
time delay on the luminosity function and redshift distribution.
We find that in this case the local rate of short GRBs is $\sim
0.8 h_{70}^3$Gpc$^{-3}$yr$^{-1}$. Assuming that all binary merging
systems lead to short GRBs, we find a typical jet opening angle of
$1.6^\circ$. }
\maketitle

\section{Introduction}

Our knowledge of gamma-ray bursts (GRBs) has improved enormously
with the detection of optical afterglows that  allowed
determination of GRBs redshifts. We know now that GRBs are
cosmological, that their emission is beamed in narrow jets, and
that the isotropic equivalent energy emitted by gamma-rays ranges
from $10^{51}$ to $10^{54}$ergs while the actual emitted energy is
around $10^{51}$ergs. Moreover, afterglow studies suggest that
GRBs are associated with the collapse of massive stars and that
the burst rate may be linked to the rate of star formation in
galaxies.

Our knowledge is limited, however, to long-duration GRBs with
$T_{90}>2$sec, since no afterglows have been observed so far for
short bursts (Gandolfi et al. 2000) with $T_{90}<2$sec (see
Kouveliotou et al. (1993) for the classification of long and short
GRBs). While current detectors are less sensitive to short bursts,
it is not clear whether this is an observational artifact or a
real feature (see e.g. Panaitescu, Kumar \& Narayan 2001).
Hopefully the upcoming detector Swift will answer this question.
As a consequence there is no direct knowledge of the redshifts,
luminosities, and space densities of short bursts.

In this paper we describe an attempt to determine the luminosity
function and formation rate of short GRBs from the BATSE peak flux
distribution. These two quantities are fundamental to
understanding the nature of these objects. Unfortunately the
observed flux distribution is a convolution of these two unknown
functions, so it is impossible to determine both functions without
additional information.  We must, therefore,  assume the formation
rate of short bursts and determine the luminosity function, or
vice versa.  Cohen and Piran (1995) have shown that the observed
BATSE flux distribution can be fitted with very different
luminosity functions depending on the choice of the GRB rate. As
it is easier to make physical assumptions about the rate of short
GRBs, we derive the luminosity function for two physical
assumptions concerning the rate of short GRBs. We, first,  assume
that short GRBs evolve in redshift like long ones and follow the
star formation rate (SFR). Then we examine the possibility that
short GRBs arise from neutron star mergers that lag behind the SFR
by a specific distribution of time lags.

We parameterize our luminosity function following Schmidt (2001)
and deterimne its parameters by obtaining the best fit to the
differential peak flux distribution. Our analysis is slightly more
elaborated than the one used by Schmidt (2001), who only fitted
the first moment of the peak flux distribution $\langle V/V_{max}
\rangle$. Given,  the uncertainties and the assumptions concerning
the shape of the luminosity function his method is, however,
sufficient enough to give an idea of the relevant parameters. We
find that under the assumption that both the long and  short
bursts follow the SFR, the local rate of the short bursts is more
than 1/2  the value of the long ones, while the number of short
GRBs detected by BATSE is $\sim 1/3$ of the long ones.

From  comparison of the value  $\langle V/V_{max}\rangle=0.29\pm
0.01$ of the long bursts (Guetta, Piran \& Waxman 2004, GPW) with
the one found in this paper for the short bursts $\langle
V/V_{max}\rangle =0.39\pm 0.02$, it is clear that the distribution
of observed  short bursts is different than the  distribution of
observed long ones (Mao, Narayan \& Piran 1994, Katz \& Canel
1996, Tavani 1998). In particular this implies that the population
of observed short bursts is closer on average than the population
of observed long ones. This does not imply, however, that the real
distribution of short GRBs is necessarily different from the real
long ones and it is  possible that both intrinsic distributions
are the same and both follow the SFR. In fact we find that we can
explain  most of the difference in $\langle V/V_{max}\rangle $
values and in the observed redshift distributions by the different
thresholds of BATSE for  detecting long and short bursts. For a
short burst one has to se a shorter (and hence noisier) temporal
window (Mao, Narayan \& Piran 1994). Specifically BATSE is usually
triggered on a 64msec window for short bursts and on 1sec for long
ones is less sensitive to short bursts by a factor of 4. {(this
ratio is intended for the peak flux averaged over the trigger time
scale)}. The slightly different typical peak luminosities and
spectra between long and short bursts have a small additional
effect on the observed redshift distributions.

An interesting question is whether short bursts could arise from
single peaks of long bursts in which the rest of the burst is
hidden by noise. In this case the assumption that short GRBs
follow the SFR-like long ones is correct:  Nakar and Piran (2002)
have shown that in most long bursts the second highest peak is
comparable in height to the first one, thus, implying that the
second peak should be above the noise like the first one. Short
bursts seem to be a different entity.  This  duration-hardness
distribution (Dezalay et al. 1996, Kouveliotou et al. 1996, Qin et
al. 2000) that clearly shows that there are no soft short bursts,
further support this hypothesis.  It is natural, therefore,  to
consider the possibility that short GRBs have different
progenitors than long ones. In particular is has been suggested
that short bursts originate from neutron star-neutron star (NS-NS)
mergers (Eichler et al. 1989).  In this case we can expect a delay
due to merging time between the SFR and the NS-NS merger rate
(Piran 1992, Ando 2004). We repeat, therefore, our previous
analysis under this assumption. We consider here the merging time
distribution implied the current data for binary neutron stars
(Champion et al. 2004). When this delay is taken into account, the
real  redshift distribution of the short bursts constitutes a
closer population than the long population and the difference in
the observed distributions of short and long bursts is not simply
the effect of the BATSE threshold.

This paper is structured in two main parts. In the first part we
derive the short GRB luminosity function (LF) and the observed
redshift distribution while assuming that short bursts follow the
star formation rate. In the second part we consider the
possibility that binary neutron star mergers are the progenitors
of short bursts and analyze the effect of the merging time delay
both on the LF and on the observed redshift distribution.

\section{Derivation of the luminosity function from the BATSE sample}

We consider   all the short GRBs ($T_{90}< 2$sec) (Kouveliotou et
al., 1993) detected while the BATSE onboard trigger  (Paciesas et
al. 1999) was set for 5.5$\sigma$ over background in at least two
detectors in the energy range 50-300keV. Among them we took bursts
for which $C_{\rm max}/C_{\rm min} \geq 1$ at the 64 ms timescale,
where
 $C_{\rm max}$ is the count rate in the second brightest illuminated detector
and $C_{\rm min}$ the minimum detectable rate. These selection
criteria for the short GRBs sample used by  Schmidt (2001) and are
compatible with those used by GPW for the long bursts. These are
the minimal conditions for a uniform sample with  clear selection
criteria. Note that BATSE has used several triggering modes, and
using bursts triggered differently would result in a nonuniform
sample. With this sample of 194 GRBs we find $\langle V/V_{\rm
max} \rangle=0.39\pm 0.02$, which is significantly different than
the value $\langle V/V_{\rm max} \rangle=0.29\pm 0.01$ found in
GPW for long GRBs.

Like GPW and Schmidt (2001), we consider a broken power law LF
with lower and  upper limits, $1/\Delta_1$ and $\Delta_2$,
respectively. The local luminosity function of GRB peak
luminosities $L$, defined as the co-moving space density of GRBs
in the interval $\log L$ to $\log L + d\log L$ is
\begin{equation}
\label{Lfun}
\Phi_o(L)=c_o
\left\{ \begin{array}{ll}
(L/L^*)^{-\alpha} &  L^*/\Delta_1 < L < L^* \\
(L/L^*)^{-\beta} & L^* < L < \Delta_2 L^*
\end{array}
\right. \;,
\end{equation}
where $c_o$ is a normalization constant so that the integral over
the luminosity function equals unity. We stress that this
luminosity function is the ``isotropic-equivalent" luminosity
function, i.e. it does not include a correction factor due to the
fact that GRBs are beamed.

For given values of the parameters $\Delta_1,\,\Delta_2$ we
determine the best fit values of the parameters  $\alpha,\beta$,
and $L^*$, as well as their allowed region at $1\sigma$ level, by
finding the least $\chi^2$. We use the cosmological parameters
$H_0 = 70~$km/sec/Mpc, $\Omega_M = 0.3$, and $\Omega_{\Lambda} =
0.7$.

The peak flux $P(L,z)$ of a GRB of peak luminosity $L$ observed at
redshift $z$ is
\begin{equation}
\label{peak}
P(L,z)=\frac{L}{4\pi D_L^2(z)}
\frac{C(E_1(1+z),E_2(1+z))}{C(E_1,E_2)}
\end{equation}
where $ D_L(z)$ is the bolometric luminosity distance and
$C(E_1,E_2)$ is the integral of the spectral energy distribution
within the observed window $E_1 = 50$keV and $E_2=300$keV. Schmidt
(2001) finds that the median value of -1.1 for the spectral photon
index in this 50-300keV band for the short bursts sample. We use
this value for a simplified k-correction.

Objects with luminosity $L$ observed by BATSE with a flux limit
$P_{\rm lim}$ are detectable to a maximum redshift $z_{\rm
max}(L,P_{\rm lim})$  that can be derived from Eq. \ref{peak}. We
consider  the  average limiting flux $P_{\rm lim}\sim 1$ ph
cm$^{-2}$ sec$^{-1}$ taken from the BATSE sample. The number of
bursts with a peak flux $>P$ is given by
\begin{eqnarray}
\nonumber
  N(>P)=\int\Phi_o(L)d\log L  \\
   \int_0^{z_{max}(L,P)} \frac{R_{GRB}(z)}{1+z}
\frac{dV(z)}{dz}dz \
\end{eqnarray}
where $R_{GRB}(z)$ is the local rate of GRBs at redshift $z$, the
factor $(1+z)^{-1}$ accounts for the cosmological time dilation,
and $dV(z)/dz$ is the comoving volume element.

\section{Short GRBs that follow the SFR}

We consider, first,  the possibility that the short GRBs follow
the SFR as do long ones. Following Schmidt (2001) we employ the
parameterization of Porciani \& Madau (2001), in particular, their
SFR model SF2
\begin{eqnarray}
\label{SFR} R_{GRB}(z)
=  \rho_0 \frac{23 \exp(3.4z)}{\exp(3.4z)+22} \,
F(z,\Omega_M,\Omega_{\Lambda}) ,
\end{eqnarray}
where $\rho_0$ is the present GRB rate and
$F(z,\Omega_M,\Omega_{\Lambda})=
[\Omega_M(1+z)^3+\Omega_k(1+z)^2+\Omega_{\Lambda}]^{1/2}/(1+z)^{3/2}$,
with $\Omega_{M,\Lambda,k}$ interpreted as the present day
cosmological parameters.

Since random errors in a cumulative distribution, $N(>P)$, are
correlated, we use  the differential distributions, $n(P)\equiv
dN/dP$, for the best fit analysis. We perform a maximum likelihood
analysis to obtain the luminosity function parameters and their 1
$\sigma$ errors, keeping $\Delta_1=30$ and $\Delta_2=10$. In the
following, we show, that the results do not change significantly
when we increase the values of $\Delta_{1,2}$. If we increase
$\Delta_1$ we add weak bursts that are practically undetectable.
On the other hand, if we increase $\Delta_2$ we add a small number
of detectable bursts that are not numerous enough to influence the
analysis.

 We find: $\alpha=0.5\pm 0.4$;
$\beta=1.5^{+0.7}_{-0.5}$ and $L^*=(4.61\pm 2.17) \times
10^{51}$erg/sec. We also find that $\beta$ and $L^*$ are
positively correlated, while $\alpha$ and $\beta$ are negatively
correlated: i.e. if we increase the absolute value of $\beta$ from
1.5 to 2, we should decrease $\alpha$ from 0.5 to 0.3. The
parameters found by Schmidt (2001) are within our 1-$\sigma$
range.

The corresponding normalization yields a local rate of short GRBs
per unit volume, $\rho_0=0.11_{-0.04}^{+0.07}$
Gpc$^{-3}$yr$^{-1}$,   where we estimated the effective full-sky
coverage of our GRB sample to be $\sim 1.8$yr. The results
obtained for the long bursts (GPW) -  $\alpha=0.1,\,\,\beta=2,\,\,
L^*=6.3\times 10^{51}$erg/sec,  and
$\rho_0=0.18$Gpc$^{-3}$yr$^{-1}$ - are within the the range of 1
$\sigma$ of the best fit values found for the short bursts.
However we should remember that k-correction in these two cases is
slightly different, as typically short bursts have harder spectra.

The total isotropic equivalent energy emitted during the burst can
be estimated as $E_{\rm iso,sh}=L^*\, T_{\rm eff,sh}$, where
$T_{\rm eff,sh}$ is an ``effective'' duration of the short burst,
which is given by the ratio of the fluence and the peak flux
(Perna, Sari \& Frail 2003, Nakar, Granot \& Guetta 2004). In
practice,  $T_{\rm eff,sh}$ varies from one burst to another. We
estimated $T_{\rm eff,sh}$ by considering both the peak fluxes (in
photons/cm$^{2}$/sec), averaged over the 64ms BATSE trigger, and
the fluences in the 50-300keV energy band. We  chose this energy
band because the peak flux energy range is 50-300 keV, coinciding
with the energy range of the nominal BATSE on-board burst trigger.
In order to convert the peak fluxes to erg/cm$^{2}$/sec we used
the spectral photon index -1.1. Meanwhile $T_{\rm eff,sh}$ is
approximated by the ratio of the fluence and  peak flux. To
estimate the typical energy emitted in the 50-300 keV energy band
we consider the average value of $T_{\rm eff,sh}$ over all short
bursts of the catalog and find $\langle T_{\rm eff,sh}\rangle
=0.32$sec, which implies $E_{\rm iso,sh}\equiv L^*\langle T_{\rm
eff,sh}\rangle= 1.5\times 10^{51}$erg.

\begin{figure}[b]
{\par\centering \resizebox*{0.85\columnwidth}{!}{\includegraphics
{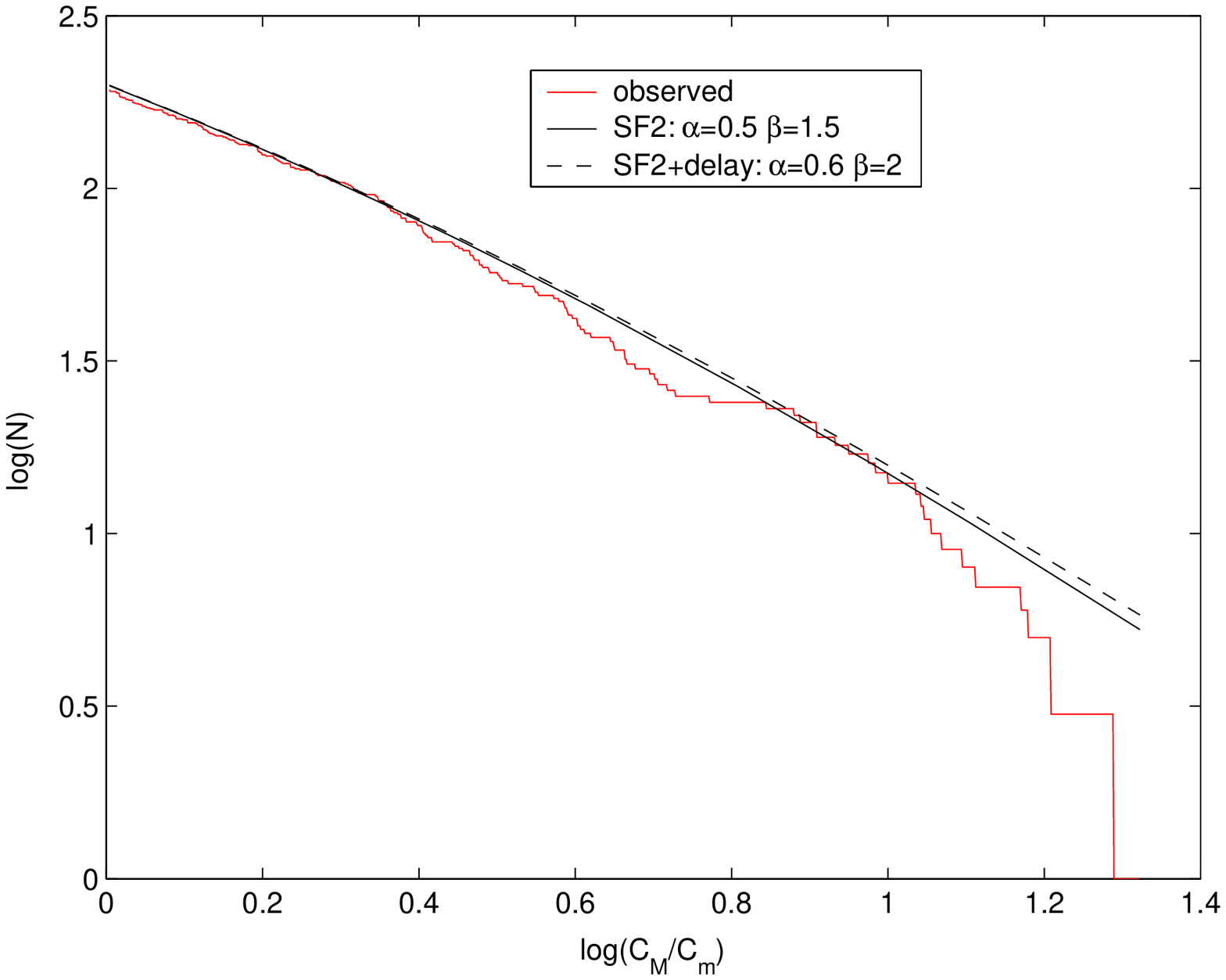}} \par}
 \caption{\label{fig1} The predicted
logN-log(P/P$_{\rm lim}$) distribution for the best fit values of
$\alpha$,$\beta$ and $L^*$ given in the text with a SF2-sfr
 and the rate obtained when the merging time is taken into account
vs. the  observed logN-log(C$_{\rm max}$/C$_{\rm min}$) taken from
the BATSE catalog. }
\end{figure}

\begin{figure}[b]
{\par\centering \resizebox*{0.85\columnwidth}{!}{\includegraphics
{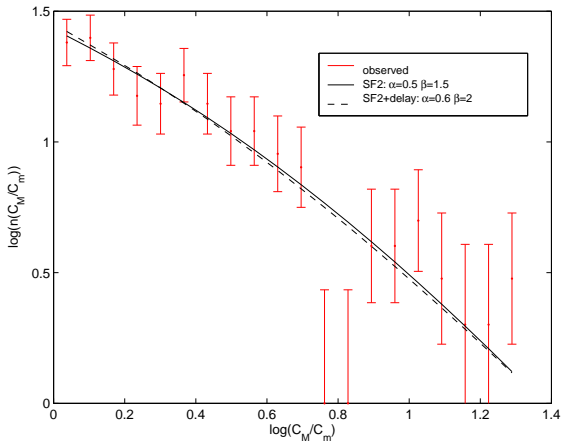}} \par}
 \caption{\label{fig2} The predicted
differential distribution, $n(P/P{\rm lim})$, for the best values
of $\alpha$, $\beta$ and $L^*$ given in the text with a SF2-sfr
and the rate obtained when the merging time is taken into account
vs. the  observed n(C$_{\rm max}$/C$_{\rm min}$) taken from the
BATSE catalog. }
\end{figure}

The total bolometric energy emitted during the burst can be
estimated by multiplying  $E_{\rm iso,sh}$ by the average value of
the ratio of the fluences in the 20-2000 keV and 50-300 keV energy
bands, $\langle f_{20-2000}/f_{50-300}\rangle \sim 6.8$. We find
$E_{\rm iso,sh,bol}=10^{52}$erg, which is a factor $\sim 20$
smaller than the corresponding value for the long bursts,  $E_{\rm
iso,l,bol}=2.1\times 10^{53}$erg, estimated in the same way (but
averaging the  peak fluxes over the 1024ms BATSE trigger, using
the spectral photon index of -1.6 and $\langle
f_{20-2000}/f_{50-300}\rangle \sim 2.7$). For the sample of long
bursts we find $\langle T_{\rm eff,l} \rangle = 12.2$sec.
Panaitescu, Kumar \& Narayan (2001) suggest that the lower total
isotropic energy of short GRB is the crucial reason their
afterglow has not been detected so far (note that the early
afterglow depends on the isotropic equivalent energy and not on
the total energy). Their assumption (following Mukherjee et al.,
1998) that the isotropic equivalent energy of short GRBs is
smaller by a factor 20 than the energy of long GRBs is consistent
with our results.

The best  fit values of the parameters  $\alpha,\beta$, and $L^*$
are  insensitive to the choice of $\Delta_1$ above a value $\sim
30$ mainly because GRBs with very low luminosity appear above the
sensitivity limit of $\sim 1$ph/cm$^2$/sec only in a very small
volume around the observer. The only parameter that changes with
$\Delta_1$ is the local rate, which clearly increases with
increasing $\Delta_1$. However, this increase is meaningless as
there is no evidence in the data that such weak bursts exist.
Increasing $\Delta_2$ also doesn't affect our results since the LF
decreases rapidly with luminosity and there are only a few bursts
with very high luminosity, regardless of the exact value of
$\Delta_2$.  Fig. 1 and 2 compare the observed integrated and
differential distributions with the predicted ones using the best
fit parameters for the luminosity function.

We can now use  the luminosity function to derive the expected
redshift distribution of the observed bursts' population in our
model. Note that the observed bursts' population depends on
properties of the detector and is different from the intrinsic
redshift distribution.
\begin{equation}
\label{redshift} N(z)= \frac{R_{GRB}(z)}{1+z} \frac{dV(z)}{dz}
\int_{L_{\rm min}(P_{\rm lim},z)}^{L_{\rm max}} \Phi_o(L)d\log L \
 ,
\end{equation}
where $L_{\rm min}(P_{\rm lim})$ is the luminosity corresponding
to minimum peak flux $P_{\rm lim}$ for a burst at redshift z and
$L_{\rm max}=L^*\times \Delta_2=10\,L^*$. This  minimal peak flux
corresponds to the gamma-ray burst detector's.  In the case of
BATSE $P_{\rm lim}\sim 1$ ph/cm$^{2}$/sec for short bursts, and
$P_{\rm lim}\sim 0.25$ph/cm$^{2}$/sec for long ones.

Fig. 3 shows the redshift distribution of long and short bursts
under different assumptions. The dot-dashed line represents  the
long GRBs redshift distribution, assuming a SF2-SFR and
considering the BATSE threshold for long bursts, $P_{\rm lim}=0.25
$ph/ cm$^2$/sec. The solid line then depicts  the short GRBs
redshift distribution assuming a SF2 SFR and considering the BATSE
threshold for short bursts $P_{\rm lim}=1 $ph/ cm$^2$/sec while
the dashed line  represents  the long GRBs redshift distribution
assuming a SF2-SFR and considering the same BATSE threshold of the
short bursts, $P_{\rm lim}=0.25 $ph/ cm$^2$/sec.

From this figure it is clear that differences  between long and
short bursts in both the $\langle V/V_{max}\rangle $ value and in
the expected observed redshift distributions arise mostly from the
differences in the threshold fluxes that trigger the BATSE
detector. In fact if we consider the triggering of long GRBs on
the 64 ms channel, thus reducing the effective sensitivity, we
find $\langle V/V_{max}\rangle \sim 0.37\pm 0.02$, which is close
to the short bursts' value. When the difference in threshold is
taken into account we find that long bursts are detected up to
higher redshifts as is evident from Fig. 3. The effect of the
different thresholds is much stronger than the differences in the
typical peak luminosities and spectra between short and long
bursts, which also contribute,  albeit at a smaller amount, to the
difference.

\begin{figure}
{\par\centering \resizebox*{0.95\columnwidth}{!}{\includegraphics
{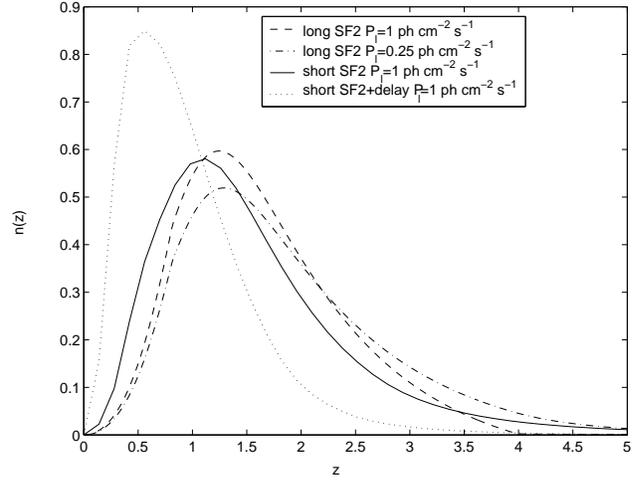}} \par} \caption{\label{fig2} Predictions for
the  observed differential redshift distributions of long bursts
(GPW) with SF2-sfr for different threshold flux values, $P_l$,
short bursts with SF2-sfr and short bursts when the delay due to
the merging time of the binary system is taken into account. }
\end{figure}

\section{NS-NS mergers as short GRBs progenitors}

There are several reasons to think that short GRBs may be linked
to binary neutron star mergers (see e.g. Narayan, Piran \& Kumar
2001). In such a case the short GRB rate is  given by the
convolution of the star formation rate with the distribution
$P_m(\tau)$ of the merging time $\tau$ of the binary system (Piran
1992, Ando 2004):
\begin{equation}
\label{rate} R_{\rm GRB}(\tau) \propto \int_{\tau_F}^{t} d\tau'
R_{\rm SF2} P_m(\tau-\tau') .
\end{equation}

The merging time distribution strongly  depends on the
distribution of the initial orbital separation $a$ between the two
stars $\tau\propto a^4$ and the distribution of initial
eccentricities. Both are unknown. Champion et al. (2004) provide a
list of known relativistic binary pulsar systems and their orbital
parameters. Six of these objects are classified as double neutron
star binaries, while the nature of a seventh pulsar system, PSR
J1829+2456, still needs to be clarified. They also list
coalescence times due to gravitational radiation emission for each
system (see Fig.4). From the coalescence time distribution it
seems that $P(\log(\tau))d\log(\tau)\sim$ const, implying
$P_m(\tau)\propto 1/\tau$, in agreement with the suggestion by
Piran (1992). Note, however, that selection effects that have to
do with the difficulty of detecting binary pulsars in close orbit,
due to the large and rapid varying Doppler shift, may
significantly affect the short time distribution below a few
hundred million years.

\begin{figure}
{\par\centering \resizebox*{0.95\columnwidth}{!}{\includegraphics
{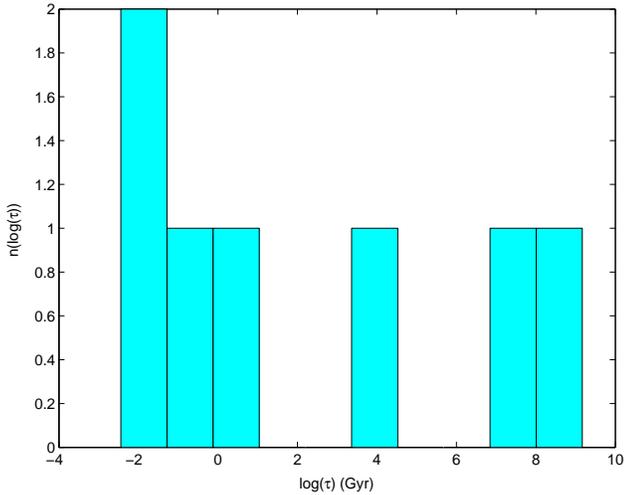}} \par} \caption{\label{fig2} The merging time
distribution implied by the data given in Champion et al. (2004).}
\end{figure}

The binaries listed in Champion et al. (2004) span a wide range of
orbital parameters but can be broadly split into two groups that
will or will not coalesce within a Hubble time, with PSR
J1829+2456 ($\tau=60$ Gyr) lying at the border between these two
groups. It is clear that the binaries not merging within a Hubble
time will not affect neither the double neutron star merger rate
calculations (Kalogera et al. 2004) nor our analysis of the
luminosity function and redshift distribution. We are aware that
the sample is too small to make predictive conclusions about
$P_m(\tau)$; on the other hand, it is the best one can do at
present.

In Fig. 5 we show the short GRB rate density as a function of
redshift due to binary neutron star mergers estimated using Eq.
(\ref{rate}). The merging time delay increases the number of both
mergers  and of  short bursts at low redshifts. This is also
evident from the differential redshift distribution depicted in
Fig. 3, which shows that the observed redshift distribution of
short GRBs is closer than the long one when the delay is taken
into account.

\begin{figure}
{\par\centering \resizebox*{0.95\columnwidth}{!}{\includegraphics
{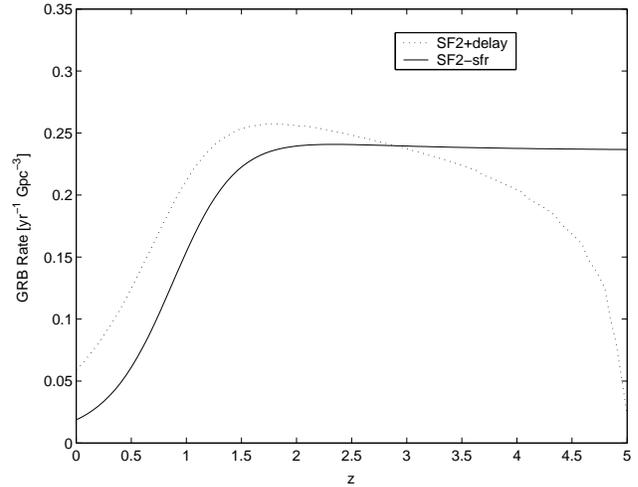}} \par} \caption{\label{fig2} Comparison of the
formation rate history of short GRBs assuming that they follow the
SF2 star formation rate like the long ones vs. the one obtained
considering their association to binary neutron star mergers}
\end{figure}

We repeated our analysis on the best fit parameters of the
luminosity function using the corrected rate and found
$\alpha=0.6, \beta=2$, and $L^*=2.2\times 10^{51}$erg/sec, which
is slightly lower than the typical luminosity found for the long
bursts and lower than the typical value found in the case where
short bursts follow the SF2-SFR. This is expected in case where
the intrinsic distribution is nearer, and hence weaker pulses are
needed. In such cases the typical isotropic-equivalent energy
$E^*_{\rm iso,sh,bol}=4.7\times 10^{51}$ erg is a factor $\sim 44$
smaller than the energy in the long bursts. The results of this
fit are shown in Figs.  1 and 2.  Assuming that short GRBs are
related to neutron star mergers and that in turn the rate of
mergers follows the SFR with a delay probability $\propto 1/\tau$,
the local rate of short GRBs is $\rho_0=0.80$/Gpc$^{3}$/yr. This
rate is a factor four larger than the local rate of long GRBs as
inferred from the analysis of GPW, and a factor of eight larger
than the value found in the case that short bursts follow the
SF2-SFR.

It is interesting to compare this rate with the observationally
inferred  rate of NS-NS mergers in our galaxy (Phinney, 1992,
Narayan, Piran \& Shemi 1992). This rate was recently reevaluated
after the discovery of PSR J1829+2456 to be rather large as
$80^{+200}_{-66}$/Myr, although the estimate contains a fair
amount of uncertainty (Kalogera et al. 2004). If we assume that
this rate is typical and that the number density of galaxies is
$\sim 10^{-2}$/Mpc$^{3}$, we find a merger rate of
$800^{+2000}_{-660}$/Gpc$^{3}$/yr. This value is significantly
larger, by a factor of $1100^{+2775}_{-900}$, than our estimate of
the local rate of short GRB, $\rho_0=0.80$/Gpc$^{3}$/yr.

If this factor arises due to beaming, it implies that short GRBs
are beamed into very narrow jets. Assuming that all the neutron
star binary mergers produce a short GRB, we can estimate the
average opening angle of the jet using this ratio to be
$1.2^\circ-4^\circ$.   The typical beaming factor we find here is
a factor 20 larger (narrower beams) than the one found for long
GRBs (GPW). The corresponding typical total energy (corrected for
beaming) of short GRBs is $\sim 2.4-20\times 10^{48}$erg, which is
smaller by 2 orders of magnitude than the total energy of long
GRBs.

These estimates are based on our estimate that $P(\tau) \propto
1/\tau$; however, those selection effects that may affect the time
distribution $P(\tau)$ are important for small $\tau$ (at 100 Myr
and lower). For this reason we also consider the extreme case in
which most binary merger on a shorter time scale and there is no
delay between  formation and merging of the binary system. In this
case if we compare the short GRB rate
$\rho_0=0.1$Gpc$^{-3}$yr$^{-1}$ with the local merger rate (even
higher under this assumption), we find a very narrow jet opening
angles $0.5^\circ-1.5^\circ$. These beaming angles imply that the
typical total energy of short GRBs is $\sim 1.3-8\times
10^{47}$erg, more than three orders of magnitude smaller than the
typical energy of long GRBs.

It is possible, of course, that every neutron star merger does not
necessarily produce a GRB or that most mergers produce weak GRBs
that are below the BATSE threshold, in which case the typical jet
opening angle and the total energy will be larger. It is also
possible that short GRBs are produced by black hole neutron star
mergers whose rate might be lower than the rate of neutron star
mergers, again leading to larger jet opening angles. It is even
possible that we currently overestimate the rate of neutron star
mergers. After all, this rate is dominated by a single object PSR
J1829+2456. Prior to its discovery the estimates of the merger
rate were lower by one order of magnitude (Kalogera 2004).

Still, it is worthwhile considering the implications of this
result, if indeed it is correct. First, we note that the narrow
beaming angles that we find, are almost the minimal allowed, as
they are comparable to the inverse of a typical Lorentz factor
$\Gamma^{-1}$ of a few hundreds. This implies, first of all, that
we should see a significant number of short burst "sideways". Such
bursts will be weaker and  have a softer spectrum, but their
duration will remain the same. A more detailed study is required
to determine if these are consistent or not with the current hard
short GRB population. Second, this implies that the energy budget
of short GRBs is much smaller than that of long GRBs. One could
easily imagine several mechanisms in which neutron star mergers
would release $\sim 10^{47}-10^{48}$ergs in gamma-rays (see
Rosswog \& Ramirez-Ruiz, 2002). Third, these narrow jet opening
angles would lead both to jet breaks taking place a few minutes
after the burst and to a rapid subsequent $t^{-2}$ decrease in the
afterglow light curve. This might be yet another reason, combined
with the low isotropic equivalent energy, for short GRB afterglows
being not detected so far.

\section{Conclusions}

This work presents the results of our derivation of the luminosity
function and the observed redshift distribution of short GRBs by
fitting the peak flux distribution taken from BATSE, and shows
that when the same intrinsic redshift distribution is assumed for
long and short bursts (SFR),  the luminosity function of short
GRBs is comparable to that of long GRBs. Their isotropic
equivalent energy is lower by a factor of 20 (SFR) or 40
(SFR+delay) due to their shorter duration. Difference in the
$\langle V/V_{\rm max}\rangle $ values and the corresponding
difference in  observed redshift distributions simply reflect the
different BATSE thresholds in the 1024msec and 64msec channels on
which long and short bursts are triggered respectively. Even if
the long and the short GRBs have slightly different typical
luminosities and spectra, the main effect on the observed redshift
distribution arises from the different threshold fluxes at which
the two populations are detected.

We  repeated our analysis assuming that short GRBs are associated
with binary neutron star mergers. In this case  a delay is
expected between the star formation rate and the merger rate. We
estimated the merging time distribution from the data of Champion
et al. (2004) (in spite of the small number of objects that are
available in this data). Naturally, the effect of the delay on the
observed redshift distribution is to increase the number of short
bursts at low redshift and to decreases their typical luminosity.
Once we compare the local rate of short GRBs obtained from our
analysis with the local coalescence rate of the binary neutron
star systems in the galaxy we find, assuming that all neutron star
binary mergers produce a short GRB,  a typical jet opening angle
that ranges between $1.2^\circ$ and $2.6^\circ$. This corresponds
to jets that are narrower, on average by a factor of $\sim 6$ (in
$\theta$) than typical jets of long GRBs. The corresponding total
energy emitted in this bursts is  $\sim 4.4 \times 10^{48}$ erg
(SFR) or  $\sim 2.4 \times 10^{47}$erg (SFR+delay), two or three
orders of magnitude below the typical energy emitted in long GRBs.
In the extreme case of no delay between the binary star formation
and the merger we obtain even narrower jets and smaller energies.

The research was supported by the RTN ``GRBs - Enigma and a
Tool" a grant from the Israeli Space Agency - SELA.

\end{document}